\documentclass[10pt]{iopart}
\usepackage{graphicx}
\usepackage[usenames]{color}

\newcommand{\difrac}[2]{\frac{\displaystyle #1}{\displaystyle #2}}

\begin{document}

\title{A Joint Search for Gravitational Wave Bursts
with AURIGA and LIGO}

\author{
L. Baggio$^{1}$, M. Bignotto$^{2}$, M. Bonaldi$^{3}$,  M. Cerdonio$^{2}$, 
M. De Rosa$^{4}$,  P. Falferi$^{3}$, S. Fattori$^{2}$,  P. Fortini$^{5}$, G. Giusfredi$^{6}$ 
M. Inguscio$^{7}$, N. Liguori$^{2}$, S. Longo$^{8}$ , F. Marin$^{7}$,  R. Mezzena$^{1}$,  
A. Mion$^{1}$, A. Ortolan$^{8}$, S.  Poggi$^{9}$, G.A. Prodi$^{1}$, F. Salemi$^{1}$, 
G. Soranzo$^{10}$, L. Taffarello$^{10}$, G. Vedovato$^{10}$, A. Vinante$^{3}$, S. Vitale$^{1}$, J.P. Zendri$^{10}$ 
}\address{The AURIGA Collaboration, http://www.auriga.lnl.infn.it}


\address{$^{1}$ Physics Department, University of Trento and INFN Gruppo 
Collegato di Trento,
Padova Section, I-38050 Povo (Trento), Italy }

\address{$^{2}$INFN\ Padova Section and Department of Physics, University of
Padova,  I-35131 Padova,  Italy}

\address{$^{3}$ Istituto di Fotonica e Nanotecnologie CNR-Fondazione Bruno Kessler and
INFN Gruppo Collegato di Trento, Padova Section, I-38050 Povo (Trento), 
Italy}

\address{$^{4}$ INOA I-80078 Pozzuoli (Napoli),  Italy and INFN Firenze 
Section, I-50121 Firenze, Italy}
\address{$^{6}$  INOA, CNR, I-50125 Arcetri, Firenze, Italy}

\address{$^{5}$ Physics Department, University of Ferrara and INFN Ferrara 
Section, I-44100 Ferrara, Italy}

\address{$^{7}$ LENS and Physics Department, University of Firenze and INFN 
Firenze Section, I-50121 Firenze, Italy}

\address{$^{8}$INFN, Laboratori Nazionali di Legnaro,  I-35020 Legnaro 
(Padova) Italy}

\address{$^{9}$ Consorzio Criospazio Ricerche, I-38050 Povo (Trento), Italy }

\address{$^{10}$ INFN Padova Section, I-35100 Padova, Italy}

\author{%
B~Abbott$^{15}$,
R~Abbott$^{15}$,
R~Adhikari$^{15}$,
J~Agresti$^{15}$,
P~Ajith$^{2}$,
B~Allen$^{2,~52}$,
R~Amin$^{19}$,
S~B~Anderson$^{15}$,
W~G~Anderson$^{52}$,
M~Arain$^{40}$
% ...
}
\address{The LIGO Scientific Collaboration, http://www.ligo.org}
\address{$^{1}$ Albert-Einstein-Institut, Max-Planck-Institut f\"ur Gravitationsphysik, D-14476 Golm, Germany}
\address{$^{2}$ Albert-Einstein-Institut, Max-Planck-Institut f\"ur Gravitationsphysik, D-30167 Hannover, Germany}
\address{$^{3}$ Andrews University, Berrien Springs, MI 49104 USA}
\address{$^{4}$ Australian National University, Canberra, 0200, Australia}
\address{$^{5}$ California Institute of Technology, Pasadena, CA  91125, USA}
\address{$^{6}$ California State University Dominguez Hills, Carson, CA  90747, USA}
\address{$^{7}$ Caltech-CaRT, Pasadena, CA  91125, USA}
\address{$^{8}$ Cardiff University, Cardiff, CF24 3AA, United Kingdom}
\address{$^{9}$ Carleton College, Northfield, MN  55057, USA}
\address{$^{10}$ Charles Sturt University, Wagga Wagga, NSW 2678, Australia}
% ...
\address{$^{15}$ LIGO - California Institute of Technology, Pasadena, CA  91125, USA}
% ...
\address{$^{19}$ Louisiana State University, Baton Rouge, LA  70803, USA}
% ...
\address{$^{40}$ University of Florida, Gainesville, FL  32611, USA}
% ...
\address{$^{52}$ University of Wisconsin-Milwaukee, Milwaukee, WI  53201, USA}
% ...

\newpage

\begin{abstract}

The first simultaneous operation of the AURIGA detector and the LIGO 
observatory was an opportunity to explore real data, 
joint analysis methods between two very different types of gravitational wave 
detectors: resonant bars and interferometers.
This paper describes a coincident gravitational wave burst search, 
where data from the LIGO interferometers are cross-correlated at the time 
of AURIGA candidate events to identify coherent transients. 
The analysis pipeline is tuned with two thresholds, on the 
signal-to-noise ratio of AURIGA candidate events and on the significance of the 
cross-correlation test in LIGO.
The false alarm rate is estimated by introducing time shifts between data sets 
and the network detection efficiency is measured with simulated signals 
with power in the narrower AURIGA band. In the absence of a  
detection,  we discuss how to set an upper limit on the rate of gravitational 
waves and to interpret it according to different source models. 
Due to the short amount of analyzed data and to the 
high rate of non-Gaussian transients in the detectors noise
at the time, the relevance of this study is 
methodological: this was the first joint search for gravitational wave bursts 
among detectors with such different spectral sensitivity and the 
first opportunity for the resonant and interferometric communities to unify 
languages and techniques in the pursuit of their common goal.
\end{abstract}

\section{Introduction}\label{s:introduction}

Gravitational wave bursts are short duration perturbations of the space-time 
metric due to such catastrophic astrophysical events as supernova core
collapses~\cite{OttBurrows06} 
or the merger and ringdown phases of binary black hole 
coalescences~\cite{FlHu:98a,FlHu:98b}. 
Over the past decade, the search for these signals has been 
independently performed by individual detectors or by homogeneous networks of 
resonant bars~\cite{IGEC} or laser interferometers~\cite{bursts-S1, 
bursts-S2,bursts-S3,bursts-S4,TAMA-only}. 
The first coincident burst analysis between interferometers
with  different broadband sensitivity and orientation was 
performed by the TAMA and LIGO Scientific  
Collaborations~\cite{LIGO-TAMA-S2}. 
That analysis required coincident detection of power excesses in
at least two LIGO interferometers and in the TAMA detector 
in the 700-2000~Hz frequency band, where all sensitivities were comparable.
The upper limit result accounted for the different antenna patterns 
with a Monte Carlo estimate of detection efficiency for sources
uniformly distributed in the sky.

This paper describes a joint burst search in a more heterogeneous
network, comprised of LIGO and AURIGA.
Although this is the first joint search with a resonant antenna,
bar data has been cross-correlated with LIGO data in the search for 
gravitational wave stochastic background, with the ALLEGRO detector.
That search provided what's, to date, the most competitive stochastic upper limit in the 905-925~Hz frequency band~\cite{LIGO-ALLEGRO}.

LIGO consists of three
 interferometers, two co-located in Hanford, WA, 
 with 2~km and 4~km baselines and one in Livingston, LA, with
a 4~km baseline, sensitive between 60 and 4000~Hz with
best performance in a 100~Hz band around 150~Hz.
AURIGA is a bar detector equipped with a capacitive resonant transducer,
located in Legnaro (PD), Italy.
In 2003 the AURIGA detector resumed data acquisition after upgrades that 
enlarged its sensitive
band to 850-950~Hz, from the $\sim$2~Hz bandwidth of the 1997-1999 
run~\cite{status-AURIGA-2003,Amaldi-Andrea,3-modes}.

Due to the different spectral shapes, an interferometer-bar coincident search 
is only sensitive to signals with power in the bar's narrower 
band. The LIGO-AURIGA analysis thus focused on short duration ($\le$20~ms) 
transients in the 850-950~Hz band, with potential target sources like 
black hole ringdowns~\cite{FlHu:98a} and binary black hole mergers~\cite{Lazarus,Lazarus2}. 

\begin{figure}[ht]
\begin{center}
\includegraphics[width=16pc]{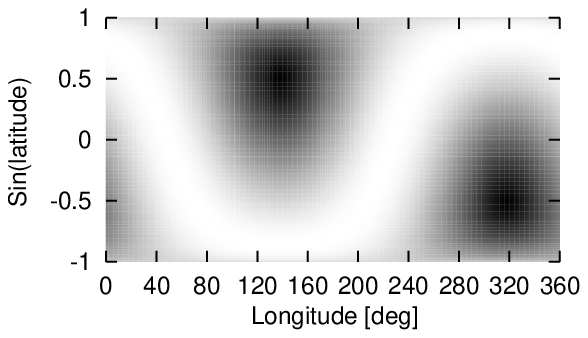}
\includegraphics[width=16pc]{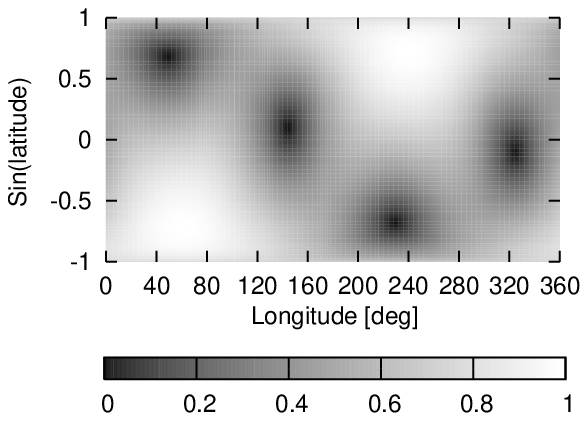}
\end{center}
\caption{\label{f:AntPattAURIGALHO} 
Antenna pattern magnitude as a function of the longitude and the sine 
of the latitude. 
Top: AURIGA; bottom:  LIGO-Hanford.}
\end{figure}

Another important difference between bars and interferometers is the sky coverage,
which depends on the detectors' shape and orientation. Figure~\ref{f:AntPattAURIGALHO}
shows the antenna pattern magnitude $\sqrt{\mbox{F}_+^2+\mbox{F}_{\times}^2}$ of the 
AURIGA and LIGO-Hanford (LHO) detectors, as a function of  
latitude and longitude.
%
Since directions of maximum LIGO sensitivity overlap with the larger 
portion of the sky visible to AURIGA, a coincident search is not
penalized by differences in antenna pattern. However, 
adding AURIGA to the detector network does not improve its overall sky coverage
either, due to the AURIGA sensitivity, which is $\sim 3$ 
times worse than LIGO~\cite{Poggi-PhDthesis,talkAURLIGOgwdaw}. 

Despite the different sensitivity and bandwidth, 
a coincident analysis between LIGO and AURIGA 
has the potential to suppress false alarms
in the LIGO network, thus increasing the confidence in the 
detection of loud signals and making 
source localization possible, with triangulation. 
Collaborative searches also increase the amount of
observation time with three or more  operating detectors.
For this reason, and to bring together the expertise of two traditions in burst
analysis, the AURIGA and LIGO Scientific Collaborations pursued 
a joint search.

The analysis described in this paper follows 
the all-sky approach described in~\cite{talkAURLIGOgwdaw}, 
where data from two or three LIGO
interferometers are cross-correlated at the time of AURIGA candidate events. 
This method was tested on data from the first AURIGA and LIGO coincident 
run, a 389~hour period between December 24, 2003 and January 9, 2004, during the
third LIGO science run S3~\cite{bursts-S3} and the first run of the upgraded 
AURIGA detector \cite{Amaldi-Andrea,3-modes}. 
Only a portion of this data was used in the joint burst search, because of
the detectors' duty factors and the selection of 
validated data segments which was independently performed by the two 
collaborations~\cite{Poggi-PhDthesis,talkAURLIGOamaldi6}. 
The effective livetime available for the  analysis was:
\begin{itemize}
\item[-] 36 hours of 4-fold coincidence between AURIGA and the three LIGO interferometers; 
\item[-] 74 hours of 3-fold coincidence between AURIGA and the two LIGO Hanford interferometers, 
when data from the LIGO Livingston detector was not available. 
\end{itemize}
Other three-detector combinations including AURIGA 
were not considered, due to the low duty factor of the LIGO Livingston interferometer
in S3. The 4-fold and 3-fold data sets were separately analyzed and the outcome was 
combined into a single result.

\begin{figure}[ht]
\includegraphics[width=3.5in]{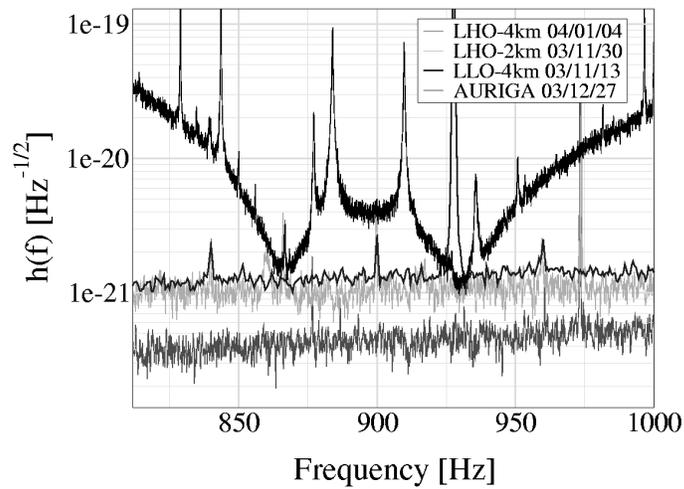}
\caption{\label{f:spectra} Best single-sided 
sensitivity spectra of AURIGA and the three LIGO interferometers during 
the joint observation. In the AURIGA spectrum, several spurious lines 
are visible while LIGO spectra present calibration lines at $973$ Hz 
for the Hanford detectors (LHO-4 km and LHO-2 km) and $927$ Hz for 
the Livingston detector (LLO-4 km).}
\end{figure}

Figure~\ref{f:spectra} shows the best single-sided 
sensitivity spectra for LIGO and AURIGA
in the 800-1000~Hz band at the time of the coincident run. 
The AURIGA spectrum contained spurious lines, due to the up-conversion 
of low frequency seismic noise. 
These lines were non stationary and could not always be filtered 
by the AURIGA data analysis; for this reason, a large portion of the data
(up to 42\%)
had to be excluded from the analysis, with 
significant impact on the livetime~\cite{Poggi-PhDthesis,talkAURLIGOamaldi6}.
The largest peak in each LIGO spectrum is a calibration line, 
filtered in the analysis.
The amount of available LIGO livetime was limited by 
several data quality factors, such as 
data acquisition problems, excessive dust at the optical tables, 
and fluctuations of the light stored in the cavities, as 
 described in~\cite{bursts-S3}.

Due to the short duration of the coincidence run 
and the non-optimal detector performances, the work
described in this paper has a methodological relevance. 
On the other hand, it is worth pointing out that 
the three-fold coincidence between AURIGA and the two Hanford 
interferometers, when Livingston was offline, allowed the
exploration of some data that would not have been searched otherwise.

\section{The analysis pipeline}
\label{s:analysis}

The joint analysis followed a statistically \textit{blind} procedure 
to avoid biases on the result: 
the  pipeline was tested, thresholds were fixed 
and procedural decisions were made before the
actual search, according to the following protocol \cite{talkAURLIGOgwdaw, talkAURLIGOamaldi6}.
\begin{enumerate}
\item AURIGA provided a list of burst candidates (triggers) in the 
validated observation time. The triggers were identified by  
matched filtering to a $\delta$-like signal, with signal-to-noise 
ratio threshold SNR$\ge 4.5$. 
Triggers at lower SNR were not included in this analysis, since 
their rate and time uncertainty increased steeply to unmanageable levels, 
with negligible improvement in detection efficiency.
Special attention was required, in this run, to address non-stationary noise
with data quality vetoes that were not needed in subsequent runs~\cite{Poggi-PhDthesis,talkAURLIGOamaldi6}.
The resulting events were auto-correlated up to about 300~s; 
this effect, particularly evident for high SNR events, 
was due to an imperfect suppression of the non-stationary spurious lines
on short time scales.
\item Data from the three LIGO interferometers at the time of 
AURIGA triggers were cross-correlated 
by the $r$-statistic waveform consistency test~\cite{rstat},
a component of the LIGO burst analysis~\cite{bursts-S2,bursts-S3} 
performed with the {\em CorrPower} code~\cite{CorrPower}. 
The test compares the 
broadband linear cross-correlation $r$ between two data streams 
to the normal distribution expected for uncorrelated data and 
computes its p-value, the probability of getting a larger $r$ 
if no correlation is present, expressed as
$\Gamma= - \mbox{log}_{10}$(p-value).
When more than two streams are involved, $\Gamma$ is the arithmetic 
mean of the values for each pair. 
The cross-correlation was performed on 20, 50 and 100~ms integration windows, 
to allow for different signal durations. 
Since the source direction was unknown, the integration windows were slid 
around each AURIGA trigger by $27 \, \mbox{ms}  +  \sigma_t$,
sum of the light travel time between AURIGA and Hanford and
of the estimated $1\sigma$ timing error of the AURIGA trigger.
The value of $\sigma_t$  depended on the SNR of each trigger, 
typically in the $5-40\,$ms range, with an average value of $17\,$ms.
The resulting $\Gamma$ was the maximum amongst all time slides and integration 
windows. Only triggers above the minimal analysis threshold of
$\Gamma\ge 4$ were considered as coincidences. 

\item A cut was applied on the
 sign of the correlation between the two Hanford interferometers,
which must be positive for a gravitational wave signal in the two co-located
detectors. 
This cut, also used in the LIGO-only analysis~\cite{bursts-S3}, 
reduced by a factor $\sim$2 the 
number of accidental coincidences, with no effect on the detection efficiency.

\item The data analysis pipeline was first 
applied, for testing purposes, to
a {\em playground} data set~\cite{playground-report}, 
which amounted to about 10\% of the livetime and was later 
excluded from the data set used in the analysis. 

\item The false alarm statistics were estimated on  
{\em off-source} data sets obtained by time shifting 
the LIGO data; more details are provided in section \ref{ss:accidental-coinc}.

\item The detection efficiency was estimated with 
Monte Carlo methods~\cite{talkAURLIGOamaldi6}, by 
simulating a population of sources uniformly distributed 
in the sky  and in polarization angle, as described
in section~\ref{ss:network-analysis}.

\item The analysis tuning consisted of setting two thresholds: 
on the SNR of the AURIGA candidate events and on the LIGO $\Gamma$ value.
 Details on the tuning procedure are available in section~\ref{ss:tuning}.

\item The statistical analysis plan was defined {\em a priori}, 
with decisions on which combination of detectors to analyze (4-fold and 3-fold)
and how to merge the results, 
the confidence level for the null hypothesis test, and the procedure 
to build the confidence belt, as described in section~\ref{ss:plan}.
 
\item Once analysis procedure and thresholds were fixed, 
the search for gravitational wave bursts was 
applied to the {\em on-source} data set.
 The statistical analysis led to confidence intervals 
 which were interpreted as rate upper limit versus amplitude curves.
 {\em A posteriori} investigations were performed on the on-source 
 results (see section~\ref{ss:comparison}), but these follow-up studies did not affect
  the statistical significance of the {\em a priori} analysis. 
\end{enumerate}

\subsection{Accidental coincidences}\label{ss:accidental-coinc}

The statistics of accidental coincidences were studied on independent
off-source data sets, obtained with unphysical time shifts
between data from the Livingston and Hanford LIGO detectors and AURIGA.  
The two Hanford detectors were not shifted relative to each other, to account 
for local Hanford correlated noise. 
The shifts applied to each LIGO site were randomly chosen 
between 7 and 100~seconds, with a minimum separation of 1~second between
shifts. 
Hanford-Livingston shifts  in the 4-detector search also
had to differ by more than 1~second.
The livetime in each shifted set varied by a few percent due to the
changing combination of data quality cuts in the various detectors. 
The net live time used in the  accidental rate estimate was
 2476.4~h from 74 shifts in the four-detector search
 and 4752.3~h from 67 shifts in the three-detector search.

\begin{figure}[ht]
\begin{center}
\includegraphics[width=14pc]{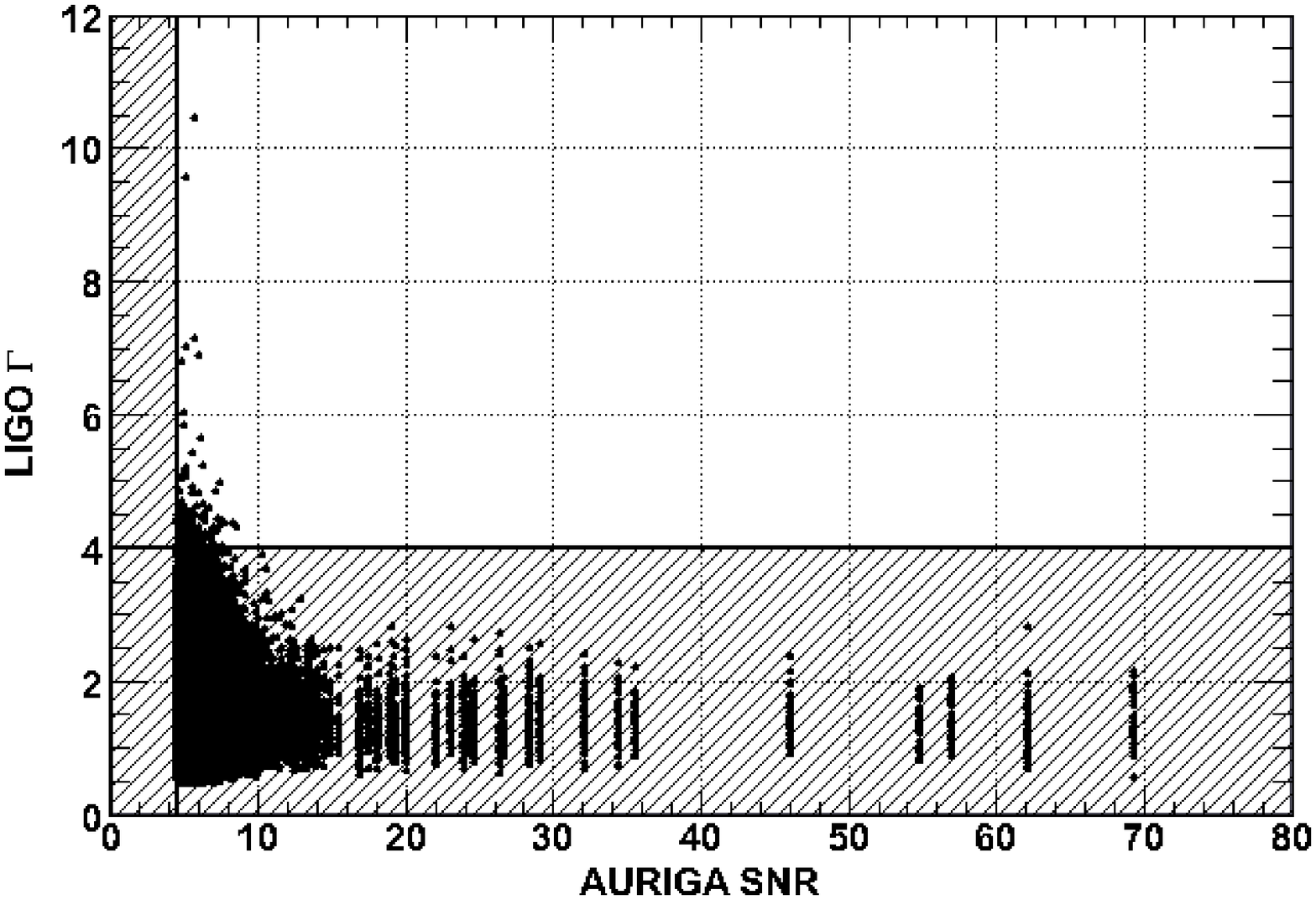}
\hspace{1pc}
\includegraphics[width=14pc]{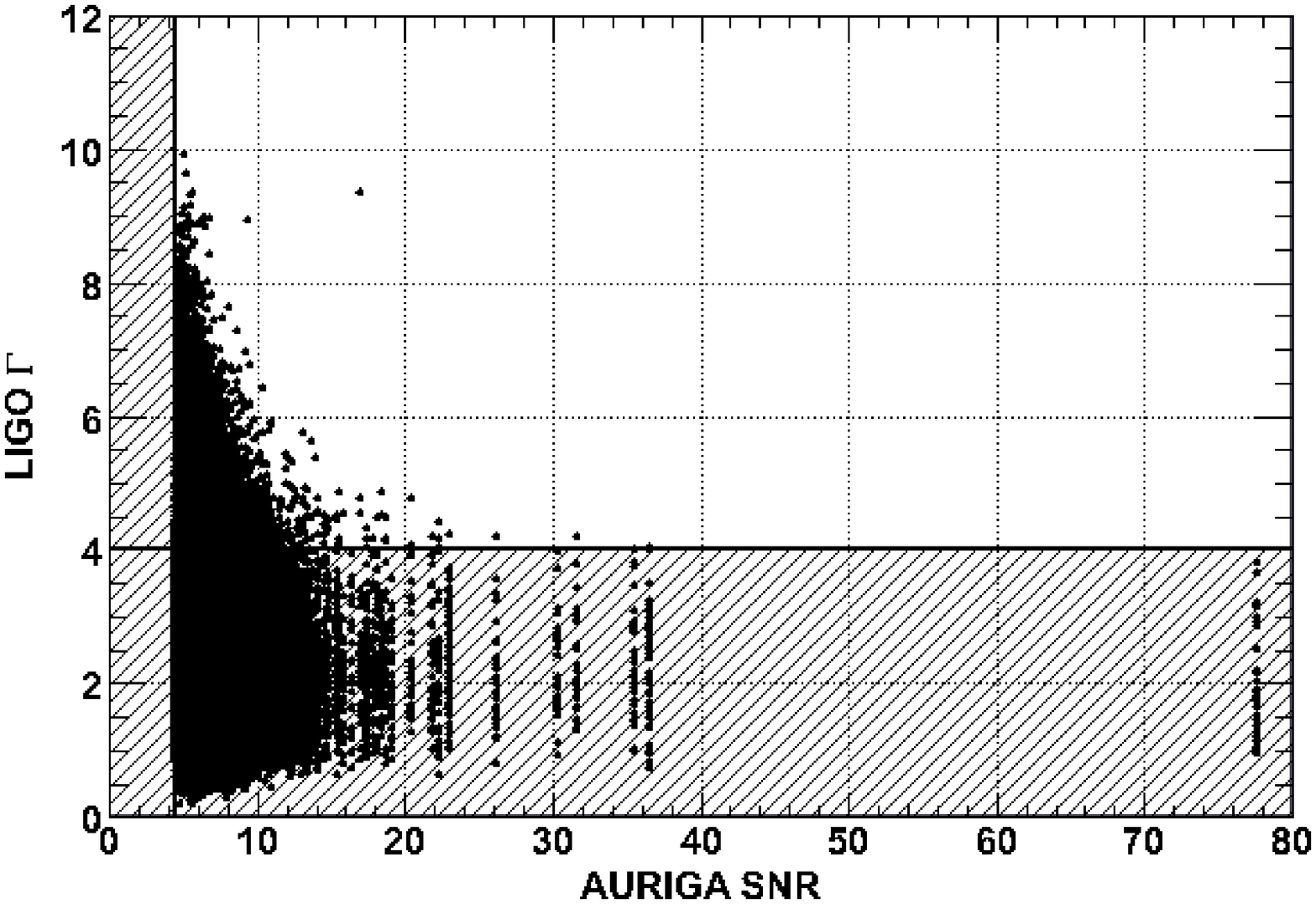}
\caption{
Scatter plots
of the LIGO $\Gamma$  versus the AURIGA SNR for 
background events surviving the cut on the sign of the
H1-H2 correlation, in the 4-detector (left) 
and in the 3-detector (right) configurations. The regions
at $\mbox{SNR} < 4.5$ and $\Gamma < 4$ are shaded, 
as they are below the minimal analysis threshold 
and were not used in the tuning.

The vertical structures at large SNR  and $\Gamma < 4$ are an artifact,
from the cross-correlation of LIGO data at different time shifts around the
same loud AURIGA event.
%
}
\label{f:Gamma-SNR-plot}
\end{center}
\end{figure}

Figure~\ref{f:Gamma-SNR-plot} shows scatter plots
of the LIGO $\Gamma$  versus the AURIGA SNR for 
background events surviving the cut on the
Hanford-Hanford correlation sign, in the 4-detector 
and in the 3-detector configurations. The regions
at $\mbox{SNR} < 4.5$ and $\Gamma < 4$ are shaded, 
as they are below the minimal analysis threshold. 

The number of off-source accidental coincidences in each time shift should be 
Poisson distributed if the time slide measurements are independent from each other. 
For quadruple and for triple coincidences, 
a $\chi^2$ test compared the measured distributions of 
the number of accidentals to the Poisson model. The test included 
accidental coincidences with 
$\Gamma\ge 4$ and $\Gamma\ge 7.5$ for 4-fold and 3-fold coincidences, respectively. 
These thresholds were  lower than what was used in the coincidence search 
(sec.~\ref{ss:tuning}), 
to ensure a sufficiently large data sample, while the AURIGA threshold remained at
$\mbox{SNR}\ge 4.5$. 
The corresponding p-values were 34\% and 6.5\%, not inconsistent with the 
Poisson model for the expected number of accidentals. 

\subsection{Network detection efficiency}
\label{ss:network-analysis}

The detection efficiency was estimated by adding
software-generated signals to real data, according to the LIGO Mock Data Challenge procedure~\cite{MDC-S2}. 
The simulation generated gravitational waves
from sources isotropically distributed 
in the sky, with  azimuthal coordinate uniform in $[0, 2\pi]$, cosine of the polar sky coordinate uniform in $[-1, 1]$ and wave polarization angle uniform in $[0, \pi]$. 
Three waveform classes were considered~\cite{Poggi-PhDthesis,talkAURLIGOamaldi6}:
\begin{enumerate}
\item\textit{Gaussians with linear polarization}:
\begin{eqnarray}
  \left\{ \begin{array}{l l} 
h_{+}(t)  =  h_{peak} \, \mbox{e}^{-(t-t_0)^2/\tau^2} \nonumber \\
h_{\times}(t)  =  0  
  \end{array} \right. 
\end{eqnarray}
with $\tau=0.2 \,$ms.
\item\textit{sine-Gaussians with linear polarization}: 
\begin{eqnarray}
  \left\{ \begin{array}{l l}
h_{+}(t)  =  h_{peak} \, \mbox{e}^{-(t-t_0)^2/\tau^2} \, \sin(2\pi f_0(t-t_0))
\nonumber \\
h_{\times}(t)  =  0 
             \end{array} \right. \,\,\,
\label{e:sine-sinusoid}
\end{eqnarray}
with $f_0 = 900 \, $Hz, $\tau = 2/f_0 = 2.2 \, $ms and $Q \equiv \sqrt{2}\pi f_0 \tau = 8.9$. 
In this analysis we also tested cosine-Gaussians (with the $\sin$ replaced by $\cos$), and found the same sensitivity as for sine-Gaussian waveforms. 

\item\textit{Damped sinusoids with circular polarization}:
 \begin{eqnarray}
h_{+}(t)
    & = &    \left\{ \begin{array}{l l}
                 h_{peak} \, \difrac{1+\cos^2{\iota}}{2} \, \cos[2\pi f_0 (t-t_0)] \, \e^{-(t-t_0)/\tau}    &  t-t_0 \ge 0 \, , \nonumber\\
                 h_{peak} \, \difrac{1+\cos^2{\iota}} {2}\, \cos[2\pi f_0 (t-t_0)] \, \e^{10(t-t_0)/\tau} &  t-t_0 < 0 \, , 
             \end{array} \right. \\
h_{\times}(t) 
    & = &    \left\{ \begin{array}{l l}
                 h_{peak} \, \cos{\iota} \,\, \sin[2\pi f_0 (t-t_0)] \, \e^{-(t-t_0)/\tau}     &  t-t_0 \ge 0 \nonumber \\
                 h_{peak} \, \cos{\iota} \,\, \sin[2\pi f_0 (t-t_0)] \, \e^{10(t-t_0)/\tau}  &  t-t_0 < 0 
             \end{array} \right.
\label{e:damped-sinusoid}
\end{eqnarray}
with $f_0=930 \, $Hz, $\tau=6 \, $ms and $cos\iota$ uniformly 
distributed in $[-1,1]$, $\iota$ being the inclination of the source with respect to the line of sight.
\end{enumerate}
Although no known astrophysical source is associated with Gaussian  and 
sine-Gaussian waveforms, they are useful because of their simple spectral 
interpretation and they are standard test waveforms in LIGO burst searches.
Damped sinusoids are closer to physical templates~\cite{FlHu:98a,Lazarus,Lazarus2}. 

The signal generation was performed with the LIGO \texttt{LDAS} software~\cite{LDAS}; 
the waveforms were added to calibrated LIGO and AURIGA data 
and the result was analyzed by the same pipeline used in the search.
For each waveform class, the simulation was repeated at different signal amplitudes to measure the 
efficiency of the network as a function of the square root of the burst energy:
$$h_{rss}  = \sqrt{\int_{-\infty}^{\infty} \!\!dt \,  \left[ h_+^2(t)+h_{\times}^2(t) \right]}
         = \sqrt{ 2 \int_{0}^{\infty} \!\!df \, {\left[ \tilde{h}_+^2(f)+\tilde{h}_{\times}^2(f) \right]}}.$$


\subsection{Analysis tuning}\label{ss:tuning}

\begin{figure}[ht]
\begin{center}
\includegraphics[width=14pc]{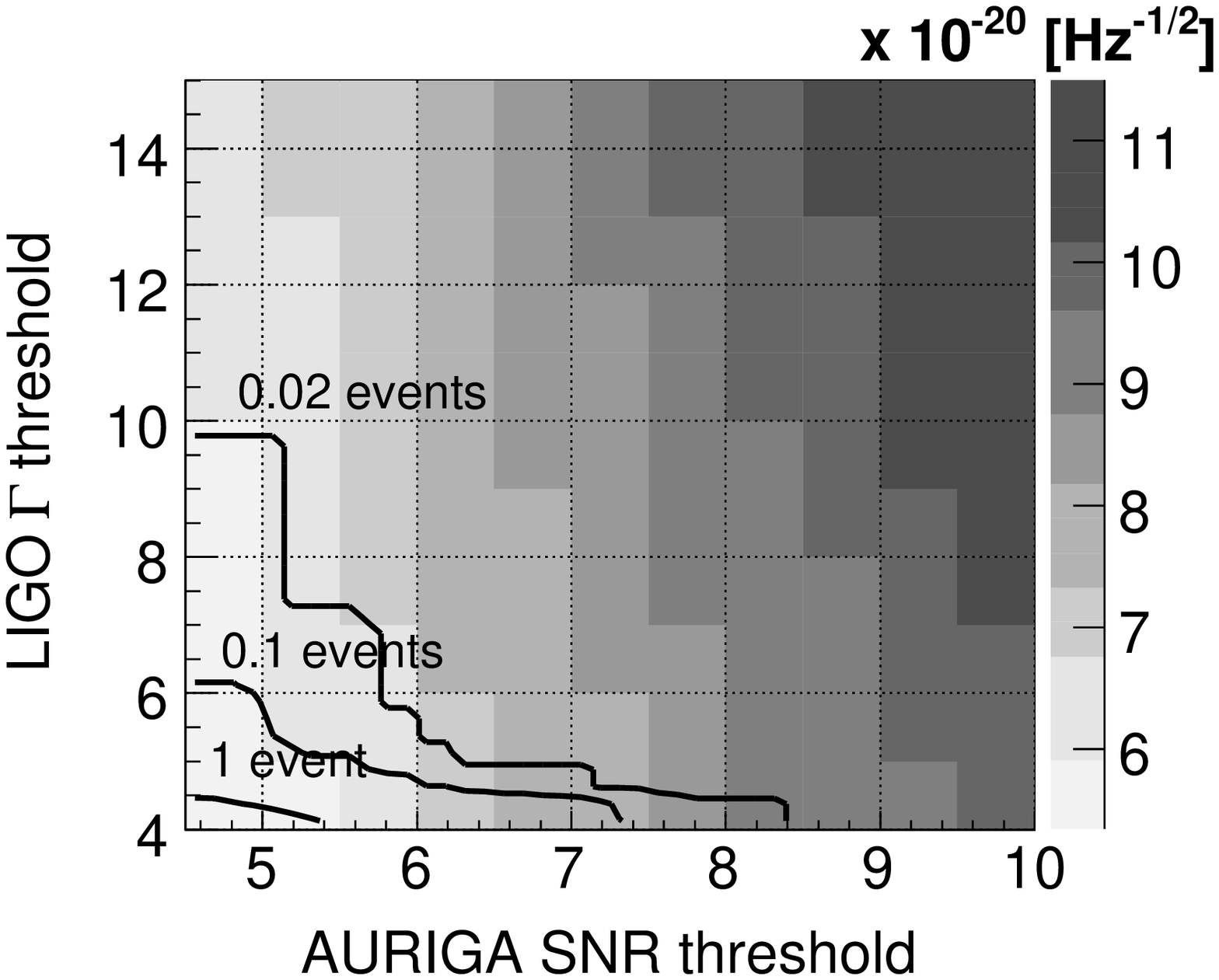}
\hspace{1pc}
\includegraphics[width=14pc]{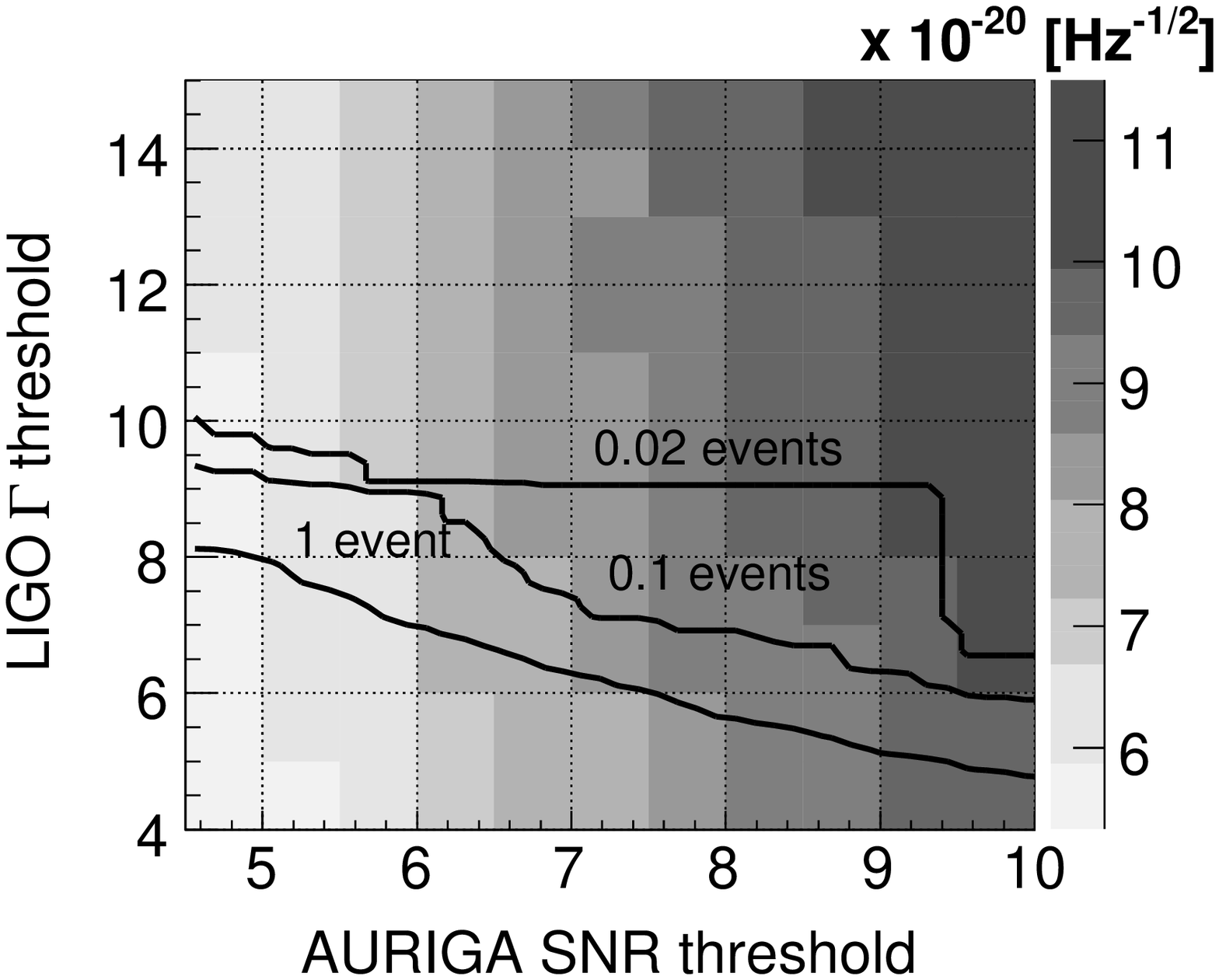}
\caption{Grayscale plot: efficiency to sine-Gaussians waveforms in terms of $h_{rss}50\%$. 
Left: 4-fold; right: 3-fold observations. 

\noindent
The contour lines indicate the number of accidental coincidences expected in the on-source data set.
}
\label{f:tuning-plot-BW}
\end{center}
\end{figure}

The analysis thresholds were chosen to 
maximize the detection efficiency with an expected number of
accidental coincidences smaller than 0.1 in each of 
the three and four detector searches.
 Figure~\ref{f:tuning-plot-BW} shows contour plots of 
the number of accidental coincidences expected in the on-source data set, 
the original un-shifted data that may include a gravitational wave signal, 
as a function of the $\Gamma$ and SNR thresholds. 
 This quantity is the number of accidental coincidences found in 
 the time shifted data, scaled by the ratio of on-source to
 off-source livetimes.
The plots also show the detectability of sine-Gaussian waveforms, expressed as 
$h_{rss}50\%$, the signal amplitude with  $50\%$ detection 
probability. 

For all tested waveforms, the detection efficiency in the 4-fold and 3-fold searches 
 are the same, within 10\%;
their value is dominated by the AURIGA sensitivity at
 $\mbox{SNR}\ge 4.5$. 
This observation, together with the shape of the 
accidental rate contour plots, indicates that the best strategy
for the suppression of accidental coincidences with minimal impact on detection efficiency is 
to increase the $\Gamma$ threshold and leave the SNR threshold at the exchange value of 4.5. 
The analysis thresholds were chosen to yield the same 
accidental rate in the two data sets: $\Gamma \ge 6$ and $\mbox{SNR} \ge 4.5$ 
for the 4-fold  search and  $\Gamma \ge 9$ and $\mbox{SNR} \ge 4.5$ for 
the 3-fold search. 

It was decided {\em a priori} to quote a single result for the entire 
observation time by merging the 4-fold and the 3-fold periods.
The number of expected accidental coincidences in the combined on-source data set is 
0.24 events in  110.0 hours, with a $1\sigma$ statistical uncertainty of $0.06$. 
Detection efficiencies with the chosen thresholds 
are listed in table~\ref{t:4fold-3fold-MDC-injections-efficiency_tuning-thresholds}.

\begin{table}[!htbp]
\begin{center}
\begin {tabular}{c|cc|cc}
\hline\hline
Waveform  & \multicolumn{2}{c|}{$h_{rss} 50\%$ [$10^{-20}\mbox{Hz}^{-1/2}$]} 
          & \multicolumn{2}{c}{$h_{rss}  90\%$ [$10^{-19}\mbox{Hz}^{-1/2}$]} \\
& 4-fold & 3-fold & 4-fold & 3-fold \\
\hline
sine-Gaussians   & 5.6 & 5.8 & 4.9 & 5.3 \\
Gaussians        & 15  &  15 & 10  &  11 \\
damped sinusoids & 5.7 & 5.7 & 3.3 & 3.4 \\
\hline\hline
\end{tabular}
\caption{Signal amplitudes with $50\%$ and $90\%$ detection efficiency 
in the 4-fold and 3-fold searches at the chosen analysis thresholds 
of $\mbox{SNR} \ge 4.5$ and $\Gamma \ge 6$ (4-fold) or $\Gamma \ge 9$ (3-fold).
These numbers do not include a $\sim 10\%$ systematic error, due to calibration 
uncertainties, and a $\sim 3\%$ $1\sigma$ statistical error, due to the small
number of simulated signals.}
\label{t:4fold-3fold-MDC-injections-efficiency_tuning-thresholds}
\end{center}
\end{table}

\subsection{Error propagation}\label{ss:error}
The detection efficiency, in a coincidence analysis, is dominated by the
least sensitive detector, in this case AURIGA. 
Monte Carlo efficiency studies show that only a small fraction of
the simulated events are lost due to the LIGO $\Gamma$ threshold; 
these events
were missed because their sky location and polarization were
in an unfavorable part of LIGO's antenna pattern. 
Since most of the simulated events were cut by the AURIGA threshold, 
the main source of systematics in this analysis
is the calibration error in AURIGA data, estimated to be $\sim 10\%$.

In addition, there is a statistical error due to the 
simulation statistics and to the uncertainty on the 
asymptotic number of injections after the veto implementation.
The $1\sigma$ statistical error on the numbers in 
table~\ref{t:4fold-3fold-MDC-injections-efficiency_tuning-thresholds} is about 3\%.

\begin{figure}[ht]
\includegraphics[width=16pc]{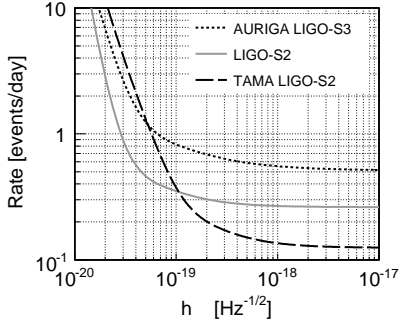}
\caption{Upper limits at $90\%$ C.L. on the gravitational wave rate 
versus $h_{rss}$ for sine-Gaussian waveforms in different 
network analyses: AURIGA-LIGO S3 (dotted line), 
LIGO-only S2 (solid line) and TAMA-LIGO S2 (dashed line). 
The sine-Gaussians have central frequency 900~Hz 
(for AURIGA-LIGO S3) and 850~Hz (for LIGO-only S2 and TAMA-LIGO S2).
 In all cases, $Q=8.9$. All sources have been modelled as uniformly 
 distributed in the sky and with random polarizations (see section~\ref{ss:network-analysis})}
\label{f:rate-hrss-comparison}
\end{figure}

Both systematic and statistical errors were taken into account in the final exclusion curve in 
figure~\ref{f:rate-hrss-comparison}.
The systematic error is propagated from the fit of the efficiency curve
to a  4-parameter sigmoid~\cite{bursts-S2,bursts-S3,bursts-S4}. 
The fit parameters were worsened to ensure a 90\% confidence 
level in the fit, following the prescriptions in~\cite{MINUIT}.
An additional, conservative shift to the left was applied to account for the 10\% error 
on the calibration uncertainty, which is the dominant error.

\subsection{Statistical interpretation plan}\label{ss:plan}

In compliance with the blind analysis approach, the statistical 
interpretation was established {\em a priori}. 
The procedure is based on a null hypothesis test to verify that the number of  
on-source coincidences is consistent with the expected distribution of 
accidentals, a Poisson with mean 0.24.
We require a $99\%$ test significance, which implies the 
null hypothesis is rejected if at least 3 
coincidences are found.

 The set of alternative hypotheses is modeled by
a Poisson distribution: 
\begin{eqnarray}
P(n|\mu)=(\mu+b)^n \exp [-(\mu+b)]/n!
\label{e:prob-Poisson}
\end{eqnarray}
where the unknown $\mu$ is the mean number of counts in excess of
 the accidental coincidences, which could be due to gravitational 
waves, to environmental couplings or to instrumental artifacts. 
Confidence intervals are established by 
the  Feldman and Cousins method with $90\%$ coverage~\cite{Feldman-Cousins}. 
Uncertainties on the estimated accidental coincidence number
are accounted for 
by taking the union of the two confidence belts with $b=0.24 \pm 3\sigma$, with $\sigma=0.06\,$. 

\begin{figure}[ht]
\includegraphics[width=16pc]{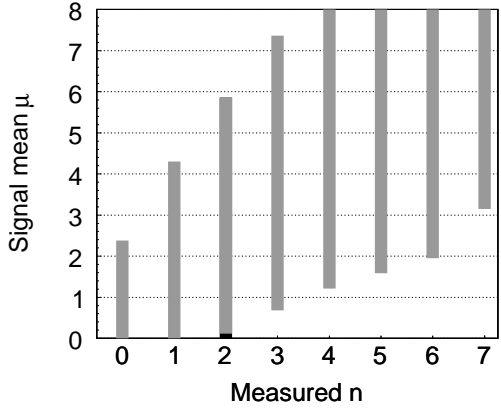}
\caption{\label{f:confidence-belt} 
Confidence belt based on the Feldman and Cousins procedure, 
with $90\%\,$ C.L., unknown Poisson signal
 mean $\mu$ and a Poisson background with $b=0.24 \pm 3\sigma$. 
The standard Feldman and Cousins construction 
 (solid lines) is modified by fixing the maximum false alarm probability 
 to $1\%$: the lower bound is fixed to 0 for 2 measured events.}
\end{figure}

The confidence belt was modified to control 
the false alarm probability according to the prescription of the null hypothesis test: 
if less than 3 events are found, and the null hypothesis is confirmed 
at 99\% C.L., we accept 
the upper bound of the Feldman and Cousins construction 
but we extend its lower bound to 0 regardless of the belt value.
The resulting confidence belt, shown in figure~\ref{f:confidence-belt}, is
slightly more conservative than the standard Feldman and Cousins belt
for small values of the signal $\mu$.
The advantage of this modification is to separate the questions 
of what is an acceptable false detection probability and what is the required
minimum coverage of the confidence intervals~\cite{gravstat}.


An excess of on-source coincidences could be due to various sources, including instrumental and environmental correlations;
the rejection of the null hypothesis or a confidence interval on $\mu$ detaching from zero do not automatically imply a gravitational wave detection.
A detection claim requires careful follow-up studies, to rule out all known sources of foreground, or independent evidence to support the astrophysical origin of the signal. 
On the other hand, an upper limit on $\mu$ can be interpreted as an upper limit on the number of GWs; therefore the upper bound of the confidence interval can be used to construct exclusion curves.


    
\section{Results}\label{s:results}



The final step consists of analyzing the on-source data sets. 
No gravitational-wave candidates were found in this search, consistent with
the  null hypothesis.
The resulting $90\%\,$CL upper limit is 2.4 events in the 
on-source data set, or 0.52~events/day in the combined 3-fold and 4-fold
data sets. 

\begin{figure}[ht]
\includegraphics[width=16pc]{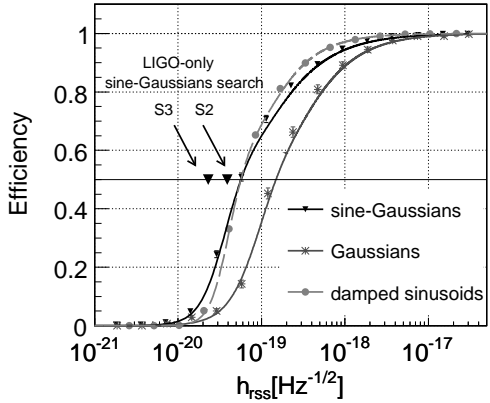}
\caption{
Average efficiency of detection (4-fold+3-fold) versus 
$h_{rss}$ for the considered waveforms: sine-Gaussians 
at $f_0=900$ Hz and $\tau=2.2$ ms, Gaussians with $\tau=0.2$ ms, 
and damped sinusoids with $f_0=930$ Hz and damping time $\tau=6$ ms. 
All sources have been modelled as uniformly distributed in 
the sky and with random polarizations 
(see section~\ref{ss:network-analysis}). 
The triangles mark the $h_{rss} 50\%$ for sine-Gaussians achieved in the
LIGO-only S2 and S3 searches.}
\label{f:upper-limit-rate-hrss}
\end{figure}

Figure~\ref{f:upper-limit-rate-hrss} shows the combined efficiency for 
this search as a function of the signal amplitude for the waveforms described 
in section~\ref{ss:network-analysis}, 
 a weighted average of the detection efficiency
of  3-fold and 4-fold searches:
\begin{equation}
\rm
\varepsilon(h_{rss})_{4fold+3fold}={{\varepsilon(h_{rss})_{4fold} \cdot T_{4fold} + \varepsilon(h_{rss})_{3fold} \cdot T_{3fold}}\over{T_{4fold}+T_{3fold}}}
\label{e:combined-efficiency}
\end{equation}
%
%
%

The $90\%\,$C.L. rate upper limit, divided by the amplitude-dependent efficiency,
yields upper limit exclusion curves similar to those obtained in 
previous searches~\cite{bursts-S2,LIGO-TAMA-S2}. 
Figure~\ref{f:rate-hrss-comparison} compares the 
sine-Gaussian exclusion curves found in this search
to those from  S2 in LIGO and LIGO-TAMA.  
The waveform used here peaks at 
900~Hz, while the previous searches used a sine-Gaussian with
850~Hz central frequency.
We verified analytically that the AURIGA detection efficiencies 
for Q=9 sine-Gaussians at 850~Hz and 900~Hz agree 
within 10\%; no significant difference is to be expected  for  the large band detectors.

The asymptotic upper limit for large amplitude signals
is inversely proportional to the observation time. 
The value for this search with $90\%$ C.L. is 0.52~events/day,
to be compared to 
0.26~events/day in the LIGO S2 search~\cite{bursts-S2} and 
0.12~events/day in the LIGO-TAMA search~\cite{LIGO-TAMA-S2}. 
The lowest asymptotic value was previously set by IGEC: 
$\sim 4 \times 10^{-3}$events/day, thanks to their longer 
observation time~\cite{IGEC}. 

The detection efficiency in this search is comparable to the LIGO-only S2
one, and a factor 2 worse than the LIGO-only S3 sensitivity. 
In the lower amplitude region, this search is an improvement 
over the IGEC search, since the AURIGA amplitude sensitivity during 
LIGO S3 was about 3 times better than the typical bar 
sensitivity in the IGEC 1997-2000 campaign (a direct comparison 
is not possible since IGEC results are not interpreted in 
terms of a source population model).
More recent data yielded significant improvements in sensitivities, by 
a factor $\sim 10$ for the LIGO S4 run~\cite{bursts-S4} and a factor 
$\sim 3$ for IGEC-2~\cite{IGEC2}.

\subsection{Diagnostics of on-source and off-source data sets}
\label{ss:comparison}



\begin{figure}[ht]
\begin{center}
\includegraphics[width=14pc]{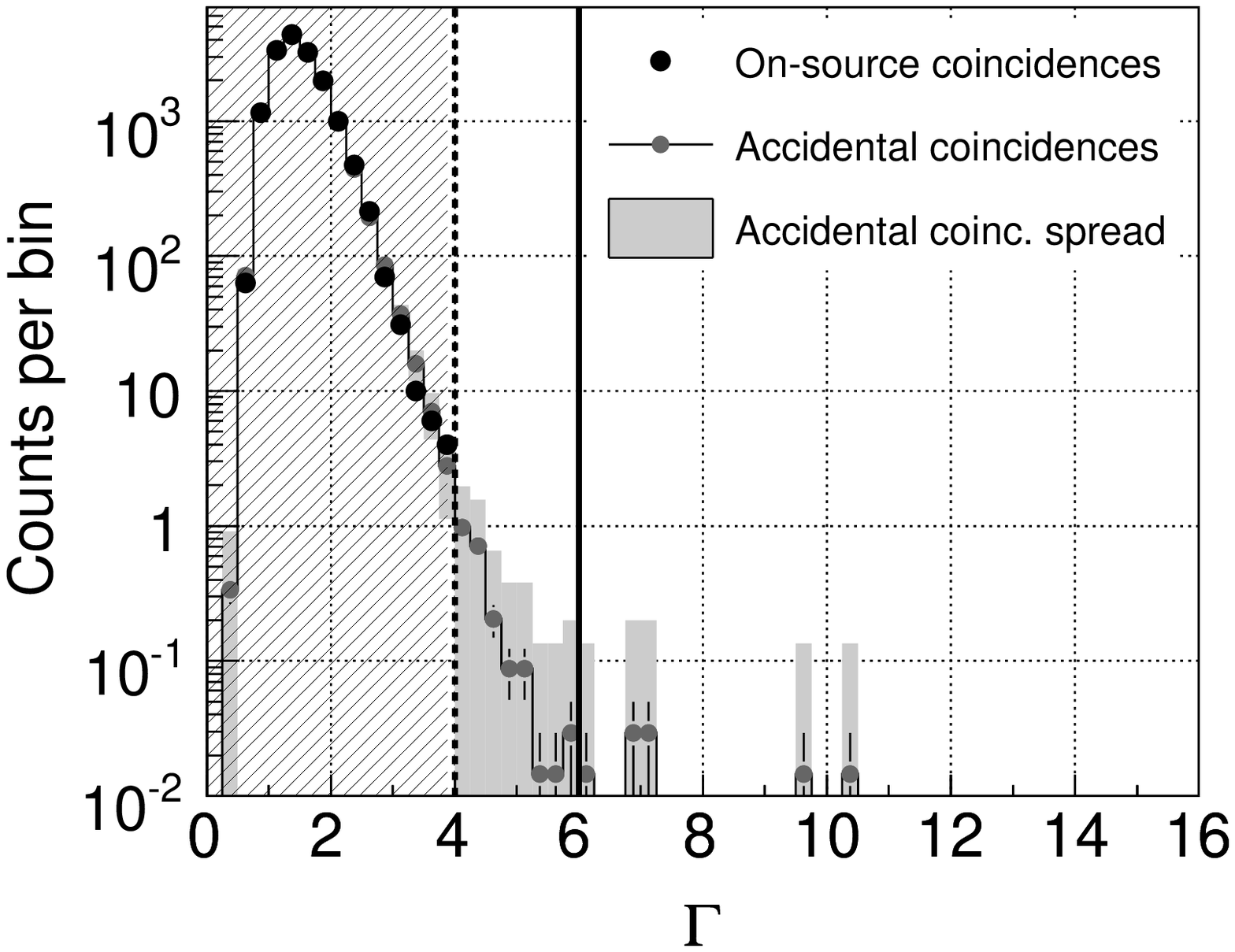}
\hspace{1pc}
\includegraphics[width=14pc]{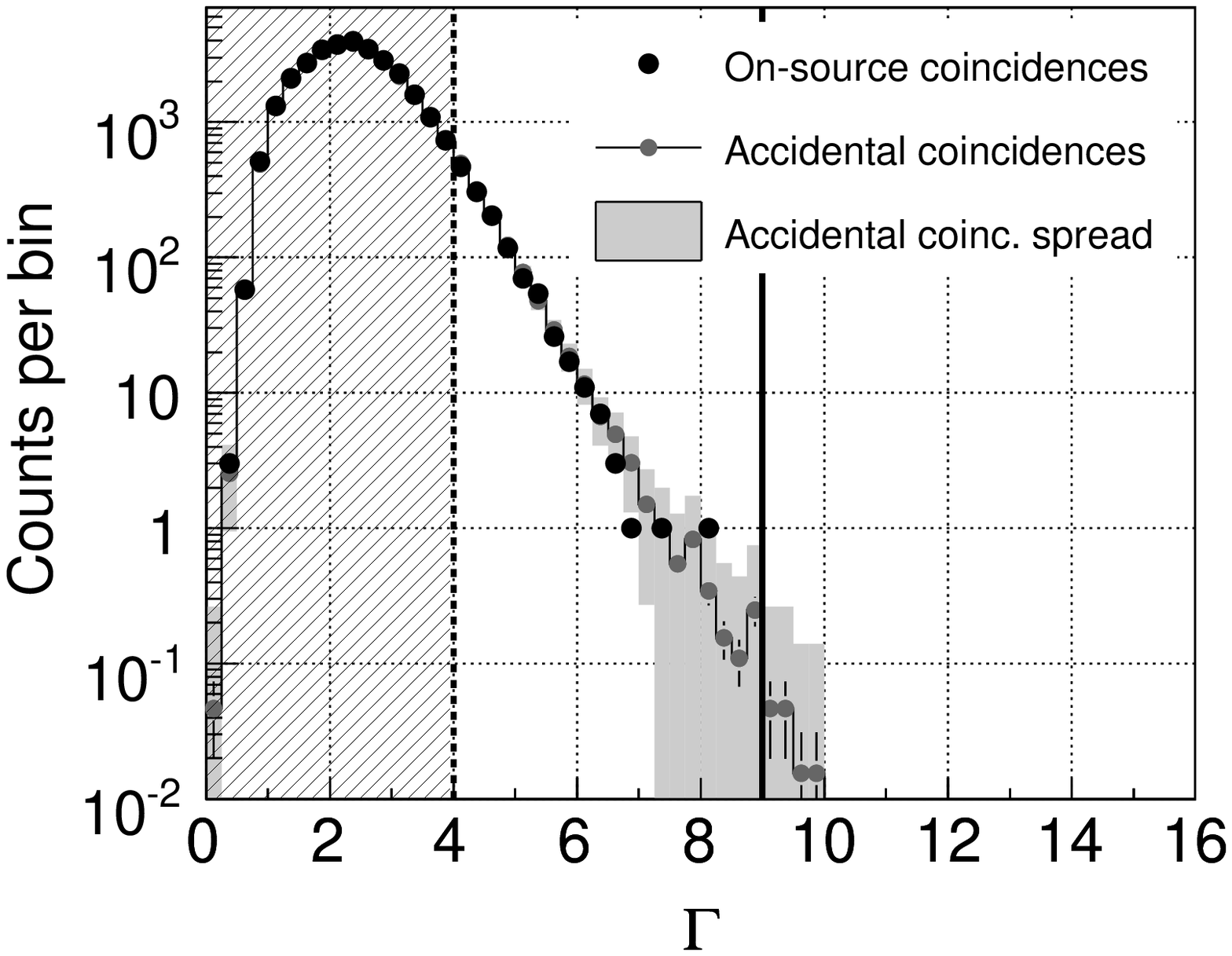}
\caption{Comparison of  $\Gamma$ distributions for four-fold (left) and 
three-fold coincidences (right). 
The black dots correspond to on-source coincidences. 
The stair-step curve with gray dots is the mean accidental background, 
estimated from off-source coincidences, with its $1\sigma$ RMS spread (shaded gray area) 
and the error on the mean (thin error bars). 
The on-source and off-source distributions are in good agreement, within the 
statistical uncertainty. 
The solid vertical lines correspond to the analysis thresholds ($\Gamma = 6$ and $\Gamma = 9$, respectively). 
The analysis was tuned only on events with $\Gamma \ge 4$ (dotted vertical lines). 
}
\label{f:gamma-background-0lag}
\end{center}
\end{figure}

The agreement between on-source and off-source coincidences was tested
comparing the $\Gamma$ distributions in Figure~\ref{f:gamma-background-0lag} 
above the minimal exchange threshold $\Gamma \ge 4$
and below the network analysis threshold ($\Gamma \ge 6$ for 4-fold and $\Gamma \ge 9$ for 3-fold).
This {\em a posteriori} test did not find a disagreement 
between  on-source and  off-source distributions. 
There were no 4-fold, on-source events with $\Gamma \ge 4$. 
For 3-fold events, the agreement between zero-lag and accidental distributions
can be confirmed with a  Kolmogorov-Smirnov test that uses
the empirical distribution of accidentals as a model, 
giving a 0.6 p-value.

In addition, we addressed the question of  whether on-source events (foreground) modified 
the distribution of accidentals (background) and biased our estimate.
This is an issue in the 3-fold AU-H1-H2 analysis where only H1 and H2 are cross-correlated 
and the measured background distribution includes instances of 
Hanford foreground events in accidental coincidence with an AURIGA shifted event.
As a result, the time shift method overestimates the number of accidentals. 
%
In this search, however, this systematic effect turned out being negligible:
the removal of all background events in accidental coincidence with 
on-source 3-fold events with $\Gamma\ge 5.5$ and SNR$\, \ge 4.5$, did not significantly
affect the $\Gamma$ histogram above threshold. 

The same question could be posed in a different way: how would the background histogram 
change if we had an actual gravitational wave event, with large $\Gamma$?
On average, the same H1-H2 event appears in $\sim 9$ background coincidences.  
A loud gravitational wave event, with $\Gamma$  above the noise, say $\Gamma=12$, 
would have appeared in the background histogram 15-20 times as a $\Gamma$ peak with a tail 
at the same Hanford time. Such an event would not have been missed, but it would have
been noticed in the tuning stages. 
The most significant consequence is that the 3-fold search is not truly blind, since a 
loud signal would easily manifest itself in the tuning data set.

\section{Conclusions}\label{ss:conclusions}

This paper describes the first joint search for gravitational wave 
bursts with a hybrid network composed of a narrow band resonant bar 
detector and broadband interferometers. 
This was a rare and valuable opportunity to bring together 
the expertise of the AURIGA and LSC collaborations and explore common 
methods on real data. The addition of the AURIGA detector to the LIGO
observatory allowed to extend the time coverage of the observations by including
also the time periods when only two of the three LIGO detectors were operating 
simultaneously with AURIGA (AU-H1-H2). This was possible thanks to the   
false alarm rate suppression contributed by AURIGA. The detection 
efficiency of this hybrid network for the tested source models was about a 
factor 2 worse than the LIGO-only efficiency, limited by the AURIGA 
detector. 
This cost however, turned out smaller than the amplitude sensitivity ratio
between AURIGA and LIGO during S3 for the same signal types (roughly, a factor 3). 

Due to the short observation time, the relevance of this study is 
methodological. The results have been interpreted in terms of source population 
models and the final upper limits are comparable to those set by LIGO alone in 
previous observations. This joint analysis followed a statistically 
{\em blind} procedure to allow an unbiased interpretation of the confidence 
of the results. In particular, the data analysis plan has been fixed {\em a 
priori} and the results are confidence intervals which ensure a minimum coverage
together with a more stringent requirement on the maximum false detection 
probability. 

Any future joint search for bursts by interferometric and resonant detectors,  
on simultaneous long-term observations, would be informed by the techniques developed
in the ground-breaking work presented in this paper.

\ack
The LIGO Scientific Collaboration (LSC) gratefully acknowledges the support of 
the United States National 
Science Foundation for the construction and operation of the LIGO Laboratory 
and the Particle Physics and Astronomy Research Council of the United Kingdom, 
the Max-Planck-Society and the State of Niedersachsen/Germany for support of 
the construction and operation of the GEO600 detector. The authors also 
gratefully acknowledge the support of the research by these agencies and by the 
Australian Research Council, the Natural Sciences and Engineering Research 
Council of Canada, the Council of Scientific and Industrial Research of India, 
the Department of Science and Technology of India, the Spanish Ministerio de 
Educacion y Ciencia, The National Aeronautics and Space Administration, 
the John Simon Guggenheim Foundation, the Alexander von Humboldt Foundation, 
the Leverhulme Trust, the David and Lucile Packard Foundation, 
the Research Corporation, and the Alfred P. Sloan Foundation.

The AURIGA Collaboration acknowledges the support of the research by the 
Istituto Nazionale di Fisica
Nucleare (INFN), the Universities of Ferrara, Firenze, Padova and Trento, the 
Center of Trento of the Istituto di Fotonica e Nanotecnologie - Istituto 
Trentino di Cultura and the Consorzio Criospazio Ricerche of Trento.

\newpage

\section*{References}

\newpage

%
%

\end{document}